# Electronic origin of the anomalous segregation behavior of Cr in Fe-rich Fe-Cr alloys

Maximilien Levesque,[1,2] Michèle Gupta,[3,*] and Raju P. Gupta[3]

[1]*École Normale Supérieure, Département de Chimie, UMR 8640 CNRS-ENS-UPMC, 24 rue Lhomond, 75005 Paris, France*
[2]*CEA, DEN, Service de Recherches de Métallurgie Physique, F-91191 Gif-sur-Yvette, France*
[3]*Laboratoire de Thermodynamique et Physico-Chimie d'Hydrures et Oxydes, Bâtiment 415, Université de Paris-Sud, 91405 Orsay, France*

The energy of segregation of Cr from the bulk to the topmost surface layer in dilute Fe-Cr alloys is endothermic but small. On the other hand, the segregation energy from the bulk to the subsurface layer is not only endothermic but relatively large. Thus, the subsurface layer presents a huge barrier, not only for the segregation of Cr from the bulk to the surface, but also for its diffusion from the surface into the bulk. This means that the topmost layer alone cannot be used to determine the barrier for Cr segregation (the segregation energy), which is determined in these alloys by the subsurface layer. We present the results of our *ab initio* electronic structure calculations on the segregation of Cr as an impurity in Fe, which shed some light on the origin of this anomalous behavior. We find that the interaction of a Cr atom situated in the subsurface layer, in contrast to those in the other layers, is quite complex with its nearest neighbor Fe atoms due to the magnetism of both Fe and Cr, leading to a highly unfavorable electronic structure. This in turn leads to a large endothermic energy of solution of Cr in the subsurface layer. These results are important for a proper understanding of the properties of the technologically important class of Fe-Cr steels with low concentrations of chromium.

## I. INTRODUCTION

The Fe-Cr ferritic steels with ∼10 atomic percent Cr are candidate materials for structural applications in the next generation of fission and fusion reactors due to their excellent mechanical properties and resistance to corrosion.[1–4] The Cr concentration of ∼10 atomic percent is found to be optimum for the ductile to brittle transition, as well as resistance to neutron-induced swelling due to defect agglomeration and helium bubble formation. This has led to considerable research and development activity, on a worldwide scale, on these alloys to reach a fundamental understanding of their properties.[5–21] This includes extensive experimental work, and associated multiscale theoretical modeling. The segregation of Cr to the surface of the alloy plays a major role in controlling many key properties, including corrosion resistance. In metals, the segregation[22] from the bulk to the surface occurs due to the difference in the solubility of the segregating element at the surface and in the bulk. If this difference is negative, meaning that the solution is more favorable (more exothermic) at the surface than in the bulk, the surface will be spontaneously covered with the segregating element, even at the lowest temperatures. This usually occurs for metalloid elements. However, in general, this is not the case for dilute alloys, which form solid solutions, since the existence of a solid solution requires the minority element to be soluble in the bulk matrix. The solution energy is, in this case, obviously less exothermic at the surface than in the bulk. In the subsurface layers, the solution energy gradually becomes more exothermic (relative to the topmost surface layer) until it reaches its value in the bulk solid. The segregation energy has thus been obtained from a straightforward difference of the solution energies at the topmost surface layer and in the bulk, and this seems to describe reasonably well the segregation behavior in most cases. The case of Fe-Cr alloys with low concentration of Cr presents an exception to this rule. In these alloys, the solution energy of Cr, both at the topmost surface layer and in the bulk, is exothermic, and the energy in the bulk is more exothermic than on the surface, as required for solid solution formation. However, the subsurface layer closest to the topmost surface layer presents, unexpectedly, a very high barrier for solubility of Cr with a highly endothermic value.[12,18–21] This situation is highly unusual in that this subsurface layer provides the main barrier for the segregation of Cr from the bulk to the surface. Further, this layer also acts as a barrier for diffusion of Cr from the surface into the bulk.

The bulk Fe-Cr alloys show a complex behavior.[5,7,12,13] At very low temperatures, the body centered cubic (bcc) solid solution of Fe-Cr shows complete miscibility at low concentrations (<10 at.% Cr). However, for higher Cr contents, there is a miscibility gap, and the solid solution phase-separates into an iron-rich bcc $\alpha$-solid solution, and a chromium-rich (>85 at.% Cr) bcc $\alpha'$-precipitate. This behavior cannot be explained in terms of a simple chemical mixing energy that does not include the magnetism of both Fe and Cr, where Cr is found to be insoluble at all concentrations in Fe. It has been shown[16–21] by *ab initio* calculations that, with the inclusion of magnetism of both Fe and Cr, the solution energy of Cr in Fe is exothermic for dilute concentrations of Cr. For higher concentrations, the solution energy becomes endothermic. This behavior is due to the competition between unfavorable Cr-Cr interactions and favorable Fe-Cr interactions.[15,17,18] The increased chromium content leads to more frustrated magnetic interactions between chromium atoms that make the solution of Cr endothermic.[17,18] At higher temperatures, the phase diagram is much simpler in so far as the bcc solid solution is concerned. It shows[9] a conventional miscibility gap with critical temperature of ∼1100 K. Several theoretical models[10–12] have successfully reproduced both the change of the sign of the mixing energy at 0 K, and the experimental phase diagram at higher temperatures.

Surface segregation is a complex phenomenon[22,23] in concentrated alloys, and unexpected behavior can be observed in certain systems, for example, NiPt alloys.[23] Further, in many magnetic materials, magnetism is known to dramatically alter the physical properties, leading to, in some cases, industrially important applications such as the INVAR effect in Fe-Ni alloys,[24] and the giant magnetoresistance in Fe-Cr alloys.[25] Thus, the inclusion of magnetism is absolutely essential for a proper understanding of the segregation phenomenon in Fe-Cr alloys. In this paper, we will not deal with concentrated Fe-Cr alloys, but rather we will investigate the interplay among local relaxations, magnetism, and electronic structure on the segregation behavior of Cr as an impurity in Fe via *ab initio* electronic structure calculations. We show, as in previous work,[12,18–21] that segregation of a chromium atom in the subsurface layer of an Fe-Cr alloy in the dilute limit is highly endothermic, in contrast to the segregation to the topmost surface layer, which is only slightly endothermic, or to the adjacent layer just below. We analyze in detail the reasons behind this anomalous behavior, and we find very subtle magnetic effects controlling this behavior. The computational details are given in Sec. II, and in Sec. III, we discuss the solution energies of Cr placed at different surface layers in Fe. This allows us to obtain the surface segregation energies that are found to be anomalous. An in-depth analysis of the role played by the surface relaxations, the antiferromagnetism of Cr, and its influence on the magnetism of surrounding Fe atoms and the electronic band structure is also presented. This analysis leads to an understanding, at an electronic and atomic scale, of this anomalous segregation behavior in terms of an indirect influence of Cr on the neighboring Fe atoms, which in turn destabilize the Cr in the subsurface layer. Concluding remarks are given in Sec. IV.

## II. COMPUTATIONAL DETAILS

The electronic structure calculations presented in this work were performed within the density functional theory (DFT) using the spin polarized version[26–28] of the Vienna *Ab Initio* Simulation Package (VASP) in the Generalized Gradient Approximation (GGA). The Projected Augmented Wave (PAW) potentials[29] were employed in conjunction with the PW91version of GGA[30] to account for exchange and correlation corrections. The 16 outer electrons of Fe (8 semi-core electrons and 8 outer valence electrons) and 12 outer electrons of Cr (6 semi-core electrons and 6 outer valence electrons) were treated as valence electrons in all computations. The semi-core states were not included in previous calculations. Sometimes, the neglect of these semi-core electrons can yield imaginary phonons for a perfectly stable crystal structure. We consider in this work the Fe (1 0 0) surface, since this surface is found to be the most stable. We used a repeated slab geometry with dimensions (2a 2a 4a) along the a, b, and c-axes, separated by a vacuum layer of 4a (∼11.4 Å) along the c-axis. The supercell thus contained 36 atoms (9 layers in the c-direction). An energy cutoff of 500 eV was employed, and the results were found to be fully converged using a (7 7 1) $k$-point grid in the Monkhorst-Pack[31] scheme. Only one Cr atom was placed in a layer. The mirror symmetry (with respect to the central layer) along the c-axis was exploited. This resulted in two Cr atoms per supercell, except for the central layer, where there was only one Cr atom, and hence only one Cr atom per supercell. The Cr-Cr distance in this geometry amounts to 2a, which is two times the next-nearest neighbor distance in Fe. The calculations were also repeated with a larger supercell with dimensions (3a 3a 4a) containing 81 atoms. In this larger cell, two types of grids were considered for the $k$-points mesh, one with (3 3 1) division and the other with a (5 5 1) division. The results in the two cases were nearly identical, indicating the full convergence of the results. The important conclusion that emerged from these calculations is that the segregation energy did not change significantly upon going from the smaller supercell with 36 atoms to the larger supercell with 81 atoms. Therefore, unless stated otherwise, the results with the smaller supercell are discussed hereafter, since the calculations with the larger supercell are computationally much more demanding. It is important to mention at this point that the calculated values of the segregation energies depend sensitively upon the size of the supercell, and this size dependence has been discussed in detail by Ponomareva *et al.*[18] We used the theoretically determined lattice constant (2.826 Å) of ferromagnetic Fe, in good agreement with the experimental value, 2.866 Å, at room temperature. This corresponds to a constant volume calculation representing the Cr concentration in the infinitely dilute limit. The atomic coordinates were fully relaxed until the total energies were converged to $10^{-4}$ eV or better, and the forces on the atoms were converged to less than $10^{-3}$ eV/Å. The magnetic moment in bulk Fe was found to be $2.20\mu_B$, i.e., in good agreement with experimental measurements, $2.22\mu_B$, and previous calculations.[13–20] For pure bulk Cr, we obtained an antiferromagnetic ground state in the bcc structure with a magnetic moment of $0.97\mu_B$ and a lattice constant of 2.866 Å. In fact, it is well known that pure Cr shows a spin density (SDW) behavior, but this SDW is quenched in the presence of impurities, and a regular antiferromagnetic character is observed. In our slab calculations, the magnetic moment at the Cr site was found to be always aligned antiferromagnetically to those of Fe atoms, and the value of this moment was found to be insensitive to the choice of the value for starting the computations. In other words, even if one decides to choose Cr to be nonmagnetic as the starting solution, the calculation readily converged to an antiferromagnetic value that was the same that would be obtained with an antiferromagnetic starting solution.

## III. RESULTS AND DISCUSSION

As already mentioned, the segregation energy is the difference of the solution energies of a solute in the final position and the initial position. The energy of solution, $E_{sol}^S$, of a single Cr atom as an impurity in Fe can be defined as

$$E_{sol}^S = \big(E_{Tot}(Fe_{N-2}Cr_2) - 2E_{Tot}(Cr) + 2E_{Tot}(Fe) - E_{Tot}(Fe_N)\big)/2,$$

where $E_{Tot}(Fe_{N-2}Cr_2)$ is the total energy of a slab containing $(N-2)$ Fe atoms and two Cr atoms symmetrically placed in surface layers S, $E_{Tot}(Fe_N)$ is the total energy of the slab without any Cr atoms, and $E_{Tot}(Fe)$ and $E_{Tot}(Cr)$ are the total energies (per atom) of pure Fe and pure Cr in their ferromagnetic and antiferromagnetic states, respectively.

TABLE I. Energies of solution, $E_{sol}$, and segregation, $E_{seg}$, of Cr in S, S-1, and S-2 layers, and in the center of the slab. The last column gives the magnetic moment, M, at the Cr site in these layers.

| Layer | $E_{sol}$ (eV) | $E_{seg}$ (eV) | M ($\mu_B$) |
|---|---|---|---|
| S | −0.085 | 0.078 | −3.1 |
| S-1 | 0.192 | 0.355 | −1.8 |
| S-2 | −0.156 | 0.007 | −1.6 |
| Central layer | −0.163 | 0.000 | −1.6 |

For the central layer, the expression is slightly different, since there is only one Cr atom per supercell. It is given by

$$E_{sol}^C = E_{Tot}(Fe_{N-1}Cr) - E_{Tot}(Cr) + E_{Tot}(Fe) - E_{Tot}(Fe_N),$$

where, as before, $E_{Tot}(Fe_{N-1}Cr)$ is the total energy of the slab containing one Cr atom at the central layer and (N − 1) Fe atoms. The segregation energy ($E_{seg}$) can then be obtained from the following relation

$$E_{seg} = E_{sol}^S - E_{sol}^C.$$

Alternatively, combining these equations, one can write

$$E_{seg} = \bigl(E_{Tot}(Fe_{N-2}Cr_2) - 2\,E_{Tot}(Fe_{N-1}Cr) + E_{Tot}(Fe_N)\bigr)/2.$$

The calculated energies of solution and segregation for different layers are given in Table I. As one can see, the solution energy is negative (exothermic) in the central layer (bulk), in the topmost surface layer (denoted S), and in the sub-subsurface layer (denoted S-2). The only exception is the subsurface layer (denoted S-1), where it is highly endothermic. This leads to an anomalously large value of the segregation energy to the S-1 layer, as shown in Fig. 1. The segregation energy is nearly constant in the bulk region, but the S-1 layer offers a large barrier for segregation, after which there is a dramatic decrease in the segregation energy. These results are in agreement with previous theoretical calculations,[12,17–21] in particular, with those where full atomic relaxations were included, as in the present work. We obtain a value of 0.090 eV for the segregation energy to the topmost (S) layer with the larger supercell containing 81 atoms, which is in reasonable agreement with the value, 0.078 eV, obtained with the smaller supercell. An important consequence of the large segregation barrier in the subsurface layer is that it also serves as a barrier for the diffusion of Cr atoms from the surface to the bulk.

Conventionally, it has been assumed that the strongest surface effects are expected to be observed at the topmost surface layer. Thus, the segregation of a Cr atom from the bulk to the surface should be controlled by the difference in the solution energies of Cr at the topmost surface layer and in the bulk. Within this assumption, with a segregation energy of 0.078 eV, obtained in the present work, the segregation of Cr should be observable at relatively low temperatures, including room temperature. The work of Suzuki et al.,[28] on the other hand, shows that this is not the case, and indeed a much higher temperature, 973 K, is required to observe segregation of Cr. With the subsurface layer providing a much stronger barrier for segregation, 0.355 eV, found in the present work, it is clear that much higher temperatures will be required, which is consistent with the work of Suzuki et al.[32]

In order to understand the origin of this anomalous behavior, we show in Fig. 2 the relaxations of interlayer distances with Cr placed at different surface layers, and we compare them with those found in pure Fe, using similar slab calculations (no Cr present in the slab). First, we observe that the relaxations of the outer surface layers, when the Cr atom is placed in the central layer, are very close to those found in pure Fe. This is expected, since the Cr atom is far from the surface, and the outer surface layers do not feel its effect. Generally speaking, the first two interlayer distances are the most affected, with the first one contracting and the second one expanding, a behavior consistent with that usually found at transition metal surfaces.[33] The strongest effect of Cr occurs when it is at the topmost surface layer (S), which leads to a contraction of the first interlayer distance by −0.068 Å, i.e., much larger than in pure Fe, ∼ − 0.02 Å. In fact, the relaxations of all the Fe atoms on the topmost surface layer are not the same but differ slightly depending upon the environment and the distance from the Cr atom. The interlayer relaxations given here are thus based on an average of the positions of the Fe atoms. It is important to emphasize that, in contrast to what one might have expected, the Cr atom does not sit exactly in the topmost surface layer containing the Fe atoms, but it is instead displaced upward by ∼0.06 Å from the plane of the Fe atoms. The contraction of the subsurface layer relative to the Cr layer is thus much smaller, only ∼0.01 Å.

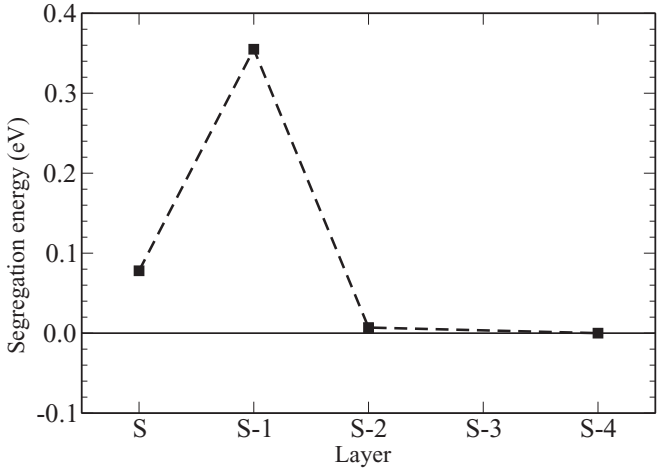

FIG. 1. Segregation energy of Cr from the bulk to different layers near the surface. The symbols S, S-1, and S-2, etc., represent, respectively, the topmost surface, the subsurface, the sub-subsurface, and other layers.

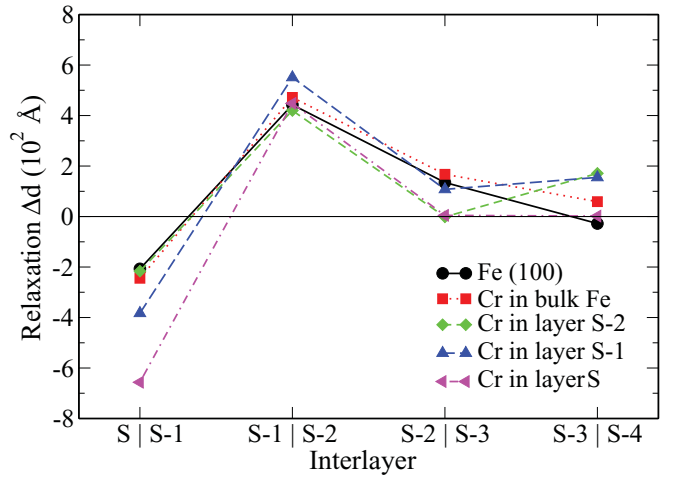

FIG. 2. (Color online) Interlayer relaxations with a Cr atom in different layers of the slab. The symbols S, S-1, and S-2 represent the topmost surface, the subsurface, and the sub-subsurface layers, respectively, and are shown in pink, blue, and green colors. The relaxations for an identical slab of pure Fe (with no Cr) are also shown, in black color, for comparison. The relaxations with the Cr atom in bulk Fe are shown in red.

This picture is somewhat different when the Cr atom is located in the subsurface layer. The contraction of the interlayer distance between the topmost (S) and the subsurface (S-1) layers containing the Fe atoms is smaller, only ∼0.038 Å. In fact, both Cr and Fe atoms in the subsurface layer are displaced upward from their ideal positions (Fe by ∼0.081 Å and Cr by ∼0.035 Å), while the Fe atoms in the topmost Fe layer are displaced upward by ∼0.043 Å from their ideal positions. The upward displacement of the Cr atom in the (S-1) layer is thus smaller than the displacement of the Fe atoms in the (S-1) layer, since the Cr atom is located at ∼0.046 Å below the Fe atoms in the (S-1) layer. This leads to a small expansion (instead of a contraction) between Cr and the topmost surface layer of Fe, of ∼0.008 Å. In both the S and (S-1) layers, the Cr atom thus moves upward relative to the ideal positions, but the displacement is much larger when it is on the topmost layer. These relaxations, in the presence of Cr, are different from those found for pure Fe, and, in general, Cr has a tendency to push further away its nearest neighbor (NN) and next-nearest neighbor (NNN) Fe atoms. As shown in the following, these relaxations modify the nature of the electronic structure considerably.

The magnetic moments of Cr with Cr in different layers are given in Table I, and they are also shown in Fig. 3. The Cr atoms are found to be aligned antiferromagnetically (shown by the negative sign in Table I) to Fe atoms in all cases. The largest magnetic moment, −3.1 $\mu_B$, is obtained at the topmost surface layer, as expected, due to the reduced coordination at the surface. These moments drop rapidly to a value of ∼−1.7 $\mu_B$ with Cr placed in the inner surface layers. Our results are in agreement with experiment[34] and previous calculations.[17–21] For pure Fe, in identical slab geometry, a magnetic moment of 2.83$\mu_B$ at the topmost surface layer is obtained, which decreases gradually to a value of 2.17$\mu_B$ at the central layer, quite close to the value, 2.20$\mu_B$, found for bulk Fe. In the presence of Cr, the magnetic moments at the Fe sites show some dispersion at the nearest and next-nearest neighbor sites in relation to Cr. The presence of a Cr atom results in a lower value of the magnetic moment at the nearest neighbor Fe sites relative to the Cr free Fe slab calculation. In moving the Cr atom from the topmost surface (S) layer with the highest magnetic moment of Cr, −3.1$\mu_B$, to the subsurface (S-1) layer, the magnetic moment at the Cr site decreases substantially (in absolute value), to −1.8$\mu_B$, and this value is not much different from the one in the sub-subsurface (S-2) layer (∼−1.6$\mu_B$) or the central layer ∼−1.74$\mu_B$. On the other hand, there is a considerable difference in the solution energies of Cr in these layers, leading to considerably different segregation energies of Cr in these layers. Hence, it is hard to correlate the behavior of the magnetic moments, either at the Fe or at the Cr sites, with the anomalous segregation behavior found for Cr segregation.

In Fig. 4, we show the total (sum of spin up and spin down) densities of states (DOS) with Cr in different surface layers and compare them with those for pure Fe in similar slab geometry. From these DOS, we see that the DOS with Cr at the subsurface (S-1) layer have been displaced slightly toward higher energies, resulting in a small peak at the Fermi level. In order to understand the origin of these changes, it is important to highlight the role of spin polarization. We therefore show in Fig. 5 the spin decomposed DOS with Cr located in various layers. On moving from the topmost surface (S) site to the subsurface (S-1) site, the magnetic moment at the Cr site changes from $-3.1\mu_B$ to $-1.8\mu_B$. Thus, the spin-up DOS is populated at the expense of the spin-down DOS at the Cr site. (Note that the sign of the magnetic moment at the Cr site is opposite to that at the Fe site.) This has an important effect on the spin decomposed DOS at the Fe sites that are nearest neighbor of Cr, as shown in Fig. 6. With Cr in the subsurface layer, the most dramatic effect is found in the DOS of the four Fe atoms in the topmost surface layer, which are the nearest neighbors (NN) of Cr. A Cr atom has four nearest neighbor Fe atoms on the topmost surface layer, and another four in the layer just below it. A comparison of the DOS at these two types of NN sites (see Fig. 6) shows that the Fe atoms on the topmost surface layer are the most affected. One can see that the spin-down DOS at these Fe sites is pushed toward higher energies, resulting in a peak at the Fermi level. A similar effect is not found in the DOS at the four NN Fe sites in the bulk just below the Cr atom. It has to be noted that, due to the presence of Cr in the subsurface layer, the magnetic moments at the Fe sites in the layers above and below have been reduced from their values in pure Fe to $2.66\mu_B$ ($2.83\mu_B$ in pure Fe) and $2.28\mu_B$ ($2.32\mu_B$ in pure Fe), but the reduction in the topmost surface layer is much larger. The magnetic moment at the NN Fe sites in the topmost surface layer is $\sim 0.4\mu_B$ higher ($2.66\mu_B$) than at NN Fe sites in the layer just below ($2.28\mu_B$), and this accounts for the lower number of electrons in the spin-down channel at the topmost Fe surface layer compared to the one just below the Cr atom. The displacement of these spin-down states at the topmost surface layer toward higher energies is clearly a destabilizing effect, and it does not favor solution of Cr in the subsurface layer. This band energy term provides a clear explanation for the lack of solubility of Cr in the subsurface layer; thus, Fe atoms on the topmost surface layer play a major role in the anomalous segregation behavior of Cr, and this cannot be understood without the inclusion of magnetism at both Fe and Cr sites.

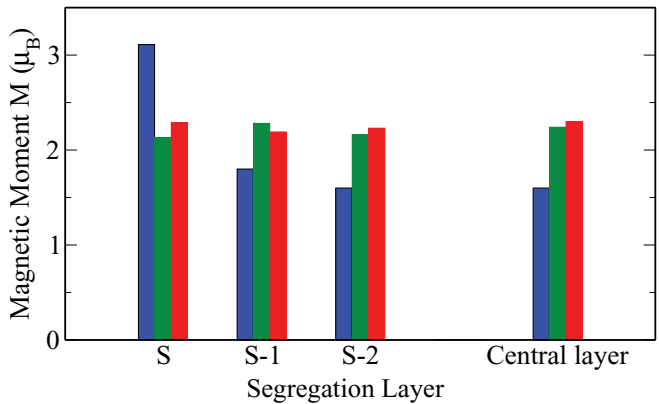

FIG. 3. (Color online) The amplitude of the magnetic moments carried by Cr in different surface layers and at its nearest neighbor Fe atoms. Note that Cr moments are aligned antiferromagnetically to those of Fe atoms. The Cr atoms are shown in blue (medium gray), nearest neighbor Fe atoms are in green (gray), and the next-nearest neighbor Fe atoms are in red (dark gray).

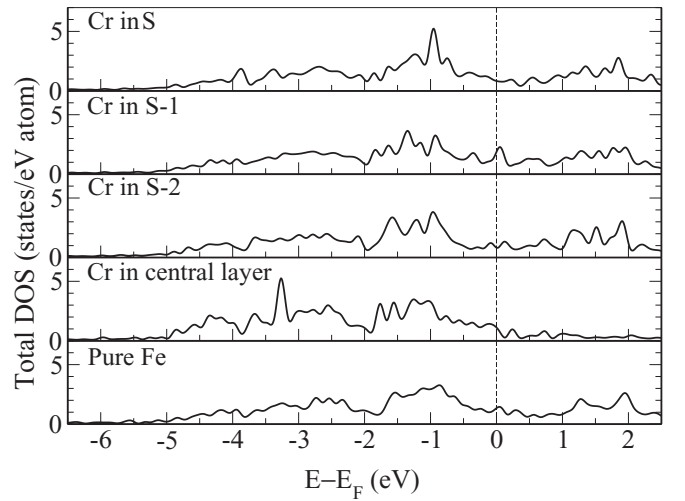

FIG. 4. The total DOS (sum of spin-up and spin-down densities of states) (in states/eV-atom) with Cr in different surface layers in the slab. Also shown for comparison are the total DOS for an identical slab of pure Fe without any Cr. The zero of the energy scale is set at the Fermi level.

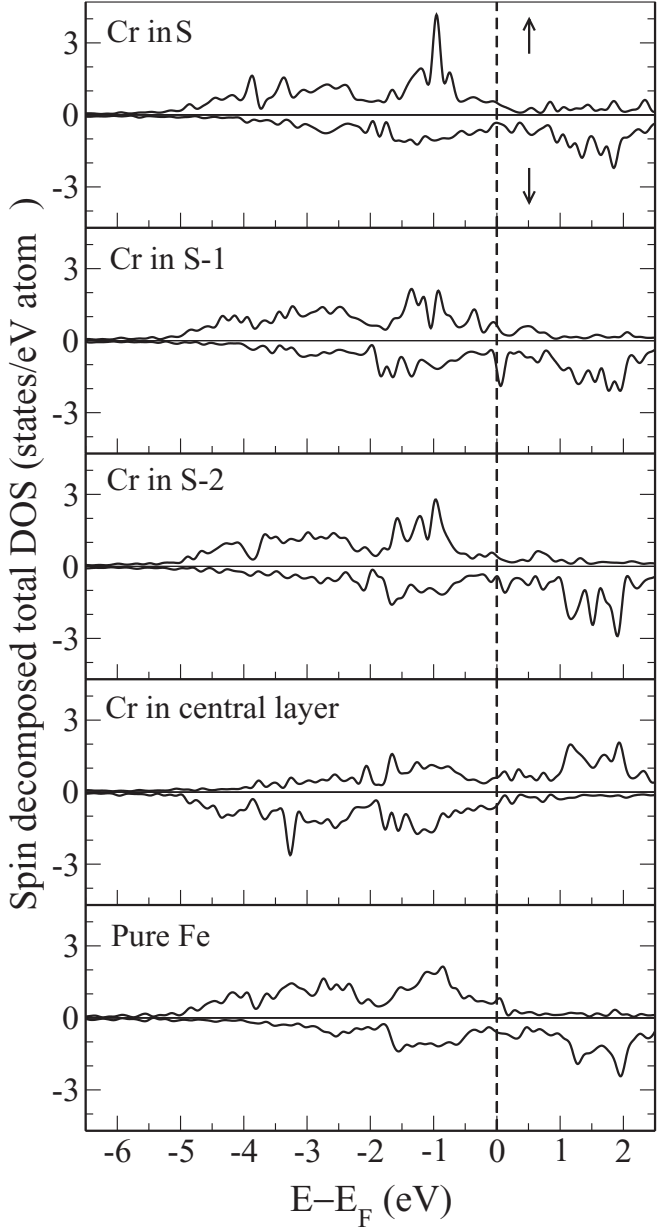

FIG. 5. Spin decomposed total DOS (in states/eV-atom) with Cr in different layers, and for a slab without any Cr. The zero of the energy scale is set at the Fermi level.

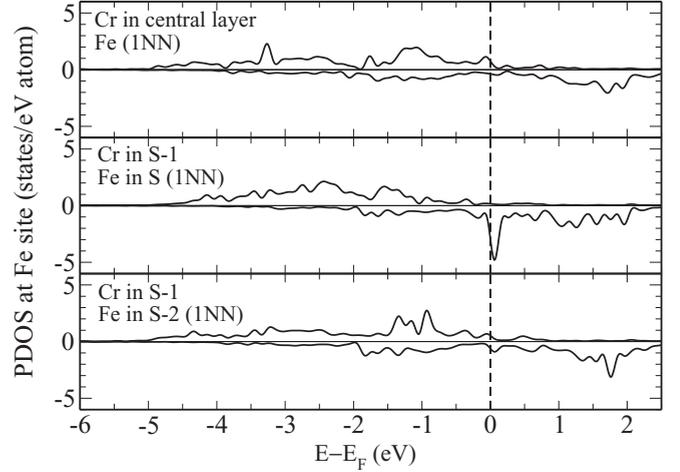

FIG. 6. Spin decomposed partial densities of states (PDOS) (in states/eV-atom) at the nearest neighbor Fe sites located in S and S-2 layers with Cr in the subsurface S-1 layer. Also shown for comparison are similar Fe site PDOS, with Cr at the central layer. The zero of energy scale is set at the Fermi level.

A detailed analysis of the atomic relaxations at the Fe sites in the presence of Cr shows that these relaxations are very similar to those found in pure Fe despite the antiferromagnetic character of Cr. However, they become substantially different when Cr is either in the surface or the subsurface layer. Our calculations show that Cr, when in the surface or the subsurface layer, exerts a repulsive type of force on the neighboring Fe atoms. When it is on the topmost surface layer, it does so by simply moving itself upward by $\sim 0.06$ Å from the topmost surface layer containing the Fe atoms. When it is in the subsurface layer, there is an expansion (instead of contraction as in pure Fe) of the interlayer separation between the layer containing Cr and the topmost surface layer of Fe atoms. This is reflected in a dramatic decrease in the magnetic moment of Cr in the subsurface layer relative to the value in the topmost layer, and a decrease in the magnetic moment of the Fe atoms at the topmost surface layer relative to the pure Fe surface, resulting in an upward shift of the spin-down states associated with the topmost Fe atoms. This is energetically a very unfavorable situation. Thus, the magnetic effects of Cr on the surface atoms do not allow the incorporation of Cr in the subsurface layer. Such destabilizing magnetic interactions are not present for Cr in the bulk, where it can be dissolved, and in the sub-subsurface layer, where the solution energy is nearly the same as in the bulk. This complicated nature of the magnetic interactions in Fe-Cr alloys makes this system complex, leading to a unique behavior of segregation of Cr in Fe.

## IV. CONCLUSIONS

The motivation of this work has been to emphasize the anomalous nature of the solution energy and the heat of segregation of Cr in the subsurface layer in Fe-rich Fe-Cr alloys, and to investigate the origin of this anomaly. In general, the segregation energy, in solid solution forming dilute alloys, increases more or less smoothly from the bulk toward the surface layers. The dilute Fe-Cr system is anomalous, since there is a large increase in the segregation energy in the layer just below the topmost layer, which then drops to a very low value at the surface. This could result in an incorrect interpretation of the data, since one usually determines the segregation energy from the difference of the solution energies in the topmost surface layer and in the bulk, and the subsurface layer is, to our knowledge, never considered. We have shown that this anomaly is of magnetic origin, and it arises due to the very complex nature of interactions between the antiferromagnetism of Cr and ferromagnetism of Fe. With Cr in the subsurface layer, the antiferromagnetism of Cr has a large influence on the electronic structure of Fe atoms on the topmost surface layer, where the electronic states in the minority channel are displaced toward higher energies, and this is not favorable for Cr solution in the subsurface layer.

## ACKNOWLEDGMENTS

We would like to thank IDRIS (Institut du Développement et des Ressources en Informatique Scientifique) for providing us access to the high-power computing (HPC) resources of GENCI (Grand Equipement National de Calcul Intensif) under project number 90189 for the work presented in this paper. M.L. thanks C.C. Fu (Commissariat à l'Energie Atomique, France) for useful discussions on DFT calculations, during his PhD Thesis on Fe-Cr alloys and surfaces.